\documentstyle[psfig]{l-aa}
\begin{document}

\thesaurus{08.02.1 --- 08.09.2 (V1101 Aql) --- 08.14.2 --- 02.01.2 --- 
Section 6}

\title{A possible orbital period for the dwarf nova V1101 Aql$^{\star}$}
\thanks{Based on observations obtained at the European Southern Observatory, 
La Silla, Chile and at the Osservatorio Astrofisico di Asiago, Italy.}

\author{Nicola Masetti\inst{1,2} \& Massimo Della Valle\inst{2}}

\institute{(1) Dipartimento di Astronomia, Universit\`a di Bologna, 
via Zamboni, 33 I-40126 Bologna, Italy\\
(2) Dipartimento di Astronomia, Universit\`a di Padova, 
vicolo dell'Osservatorio, 5 I-35122 Padua, Italy}

\offprints{N. Masetti, Univ. Bologna, masetti@astbo3.bo.astro.it}

\date{Received June 1997; accepted 7 October 1997}

\maketitle
\markboth{N. Masetti \& M. Della Valle: The period of V1101 Aql}{}

\begin{abstract}

We have performed a Discrete Fourier Transform on 136 
CCD $B$, $V$ and $R$ frames of the Z Cam--type dwarf nova V1101 Aql. 
Our analysis indicates as possible orbital period $P_{\rm orb}$ =
3$^{\rm h}$.46, though we cannot exclude the alias at 4$^{\rm h}$.00.
We estimate the distance to the system to be about 300 pc. We possibly 
discovered a bright bow--shaped nebulosity around the object.

\keywords{Binaries: close --- Stars: individual (V1101 Aql) --- Novae, 
Cataclysmic Variables --- Accretion, accretion disks}

\end{abstract}

\section{Introduction}

Dwarf Novae (hereafter DNe) are Cataclysmic Variables (CVs) characterized by
periodic outbursts, lasting few days, followed by longer periods of quiescence 
lasting from few weeks to several months. These outbursts (see Cannizzo 1993 
and references therein) are originated by cyclic instabilities in the accretion
disk surrounding a hot white dwarf (WD) or subdwarf. 
Z Cam stars form a subclass of DNe in the sense that they are characterized by 
outburst `standstills' (see Warner 1995), i.e. by prolonged phases in which 
their luminosity is halfway between maximum and quiescence. According to Osaki 
(1996), in these systems the mass transfer rate from the secondary is very 
close to the limit at which the accretion rate from the secondary ceases to 
trigger periodic instability episodes in the accretion disk and maintains the 
disk in a stable state (Frank et al. 1992). According to the observations, 
Z Cam stars are above the Period Gap (Warner 1995).

\bigskip
V1101 Aql is listed in the {\it General Catalog of Variable Stars} (Kholopov 
1987) as an irregular variable, although Richter (1961) stated that this object
was an RR Lyr star. On the contrary, Meinunger (1965), Vogt \& Bateson (1982) 
and Downes \& Shara (1993) classified it as a Z Cam-type DN. This would be 
confirmed by the observations of Pastukhova \& Shugarov (1994), which noticed 
that the star shows an ultraviolet excess, generally varies in the $B$ band 
between magnitudes 13.8 and 14.8 and has `Algol--like fadings' down to 
magnitude $B=17.3$ (these 
might possibly be observations made during the quiescent phase). Actually, they
classified V1101 Aql as a CV. The same conclusion was reached by Downes et al. 
(1995) from the analysis of spectra acquired on September 1992 and on August 
1994, when the object was at $V=14.7$ and $V=14.3$, respectively. However, it 
should be noted that the latter authors find some spectral similarities between
V1101 Aql and the class of Herbig Ae/Be stars.

In this paper we present high--time resolution photometry of V1101 Aql obtained
on September 1993 and on July 1996, together with a spectrum secured on June 
1996.
Section 2 will describe the employed instruments and the reduction techniques,
Sect. 3 will analyse the spectrophotometric data and Sect. 4 will discuss the
results. Finally, Sect. 5 will draw our conclusions.

\section{Observations}

The images of V1101 Aql are divided into two data sets. The first one has been
acquired on September 10, 1993 with the 0.9m Dutch telescope at La Silla. 
This run was composed of 71 $V$ and 3 $B$ frames.
The second data set was obtained on July 13 and 14, 1996 with the 1.2m 
telescope of the Asiago Astrophysical Observatory: here, 58 $V$, 2 $B$ and 2 
$R$ frames were collected. Table 1 reports the log of the observations.

\begin{table*}
\caption[]{Journal of the observations reported in this paper}
\begin{center}
\begin{tabular}{ccccc}
\noalign{\smallskip}
\hline
\noalign{\smallskip}
\multicolumn{5}{c}{Imaging} \\
\noalign{\smallskip}
\hline
\noalign{\smallskip}
Date & Telescope & Filter & Number & Exp. times \\
 & & or passband & of frames & (minutes) \\
\noalign{\smallskip}
\hline
\noalign{\smallskip}
\multicolumn{1}{l}{Sep. 10, 1993} & ESO 0.9m Dutch & $B,V$ & 3,71 & 2,3\\
\multicolumn{1}{l}{Jul. 13, 1996} & Asiago 1.20m & $B,V,R$ & 2,32,2 & 3 to 10\\
\multicolumn{1}{l}{Jul. 14, 1996} & Asiago 1.20m & $V$ & 26 & 5 to 10\\
\noalign{\smallskip}
\hline
\noalign{\smallskip}
\multicolumn{5}{c}{Spectroscopy} \\
\noalign{\smallskip}
\hline
\noalign{\smallskip}
\multicolumn{1}{l}{Jun. 9, 1996} & Asiago 1.80m & 3500--6000 & 1 & 30 \\
\noalign{\smallskip}
\hline
\end{tabular}
\end{center}
\end{table*}

After the standard cleaning procedure for bias and flat field, the 
frames were processed with DAOPHOT II (Stetson 1987) and {\sl ALLSTAR} inside 
MIDAS. Magnitude calibration was performed using the secondary 
photometric sequence (stars 2, 5, 6 and 7) established by Misselt (1996).
The typical error on photometry is $\la \pm$0.02 mag.
The internal magnitude differences of the comparison stars were nearly constant
within the errors (see Fig. 3a,b): this confirms that only V1101 Aql is 
responsible for the observed variability. The observation times at 
mid--exposure were then converted to Heliocentric Julian Days for each 
magnitude measurement.

\bigskip
A single spectrum of the system was taken on June 9, 1996 with the 1.80m
telescope of the Asiago Observatory, equipped with a 300 grooves mm$^{-1}$ 
grating, (wavelength range: 3500--6000 \AA), and a slit width of 2" 
(resolution of 4.3 \AA/pixel).
This spectrum was extracted and reduced with IRAF, and calibrated in wavelength
with a Fe--Ar lamp. Flux calibration was performed with the spectroscopic 
standard Feige 92.

\section{Data analysis}

\subsection{Photometry}

On September 10, 1993, the mean $V$ magnitude of the star was 14.49, whereas on
July 1996, it was $<$$V$$>=14.72$. The $B-V$ color index remained constant,
being 0.28 on September 1993 and 0.29 on July 1996; the $V-R$ was 0.16 on the 
night of July 13, 1996.

The search for periodic light variations in the $V$ band has been performed 
with a Discrete Fourier Transform (DFT) algorithm. 

To reduce the noise in the DFT power spectrum, 
we have shifted the data points to a common magnitude level, computed 
by using the mean magnitude for each night.
This could appear somewhat arbitrary, but is suggested by the fact that we are 
able to determine the mean luminosity of V1101 Aql in both parts of the 
observing run since we observed the object for more than one orbital cycle 
(see below) during each night.

\begin{figure}
\psfig{file=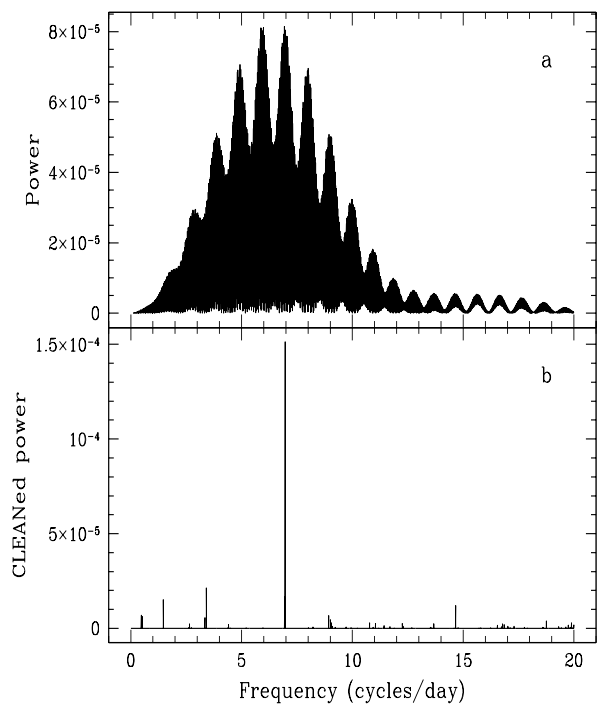,height=11cm,width=8.5cm,angle=0}
\caption[]{{\bf a} DFT and {\bf b} CLEANed power spectra of the $V$ data set.
The strongest peak in the lower panel corresponds to 0.1442 days 
(=3$^{\rm h}$.46)}
\end{figure}

The DFT power spectrum of the $V$ data points (Fig. 1a) shows a series of
peaks, the most prominent being at $\sim$7 cycles day$^{-1}$ (=0.144193
days, or 3$^{\rm h}$.46), with a slight prevalence on its one--day alias at
$\sim$6 cycles day$^{-1}$ (=0.166809 days, or 4$^{\rm h}$.00). All the peaks
belong to the same family of one--day aliases and are therefore produced by the
sampling of one single real periodicity of the $V$ lightcurve.
We therefore applied the CLEAN algorithm (Roberts et al. 1987) to discriminate
the modulation responsible for producing the alias series. The results of this
analysis, reported in Fig. 1b, suggest as real period the 3$^{\rm h}$.46 
modulation. As a further check, we subtracted this periodicity
with appropriate amplitude and phase to the $V$ data set and computed a new DFT
(Fig. 2a): its power spectrum peaks at $\sim$7 cycles day$^{-1}$, but the power
of the peak is lower than that of Fig. 1a. We followed the same procedure using 
the 4$^{\rm h}$.00 modulation (Fig. 2b): in this case, the DFT power spectrum 
is the same of Fig. 1a but the peaks had their spectral power doubled. 
Finally, we applied two least squares
best--fit methods to the $V$ data points, i.e. the Sterken's (1977)
and the Sch\"oneich--Lange's (1981) algorithms, and both methods indicate as
best fit the periodicity at 0.144 days.

The $V$ lightcurves folded with the 3$^{\rm h}$.46 and 4$^{\rm h}$.00 are 
shown in Fig. 3a,b, respectively.

\begin{figure}
\psfig{file=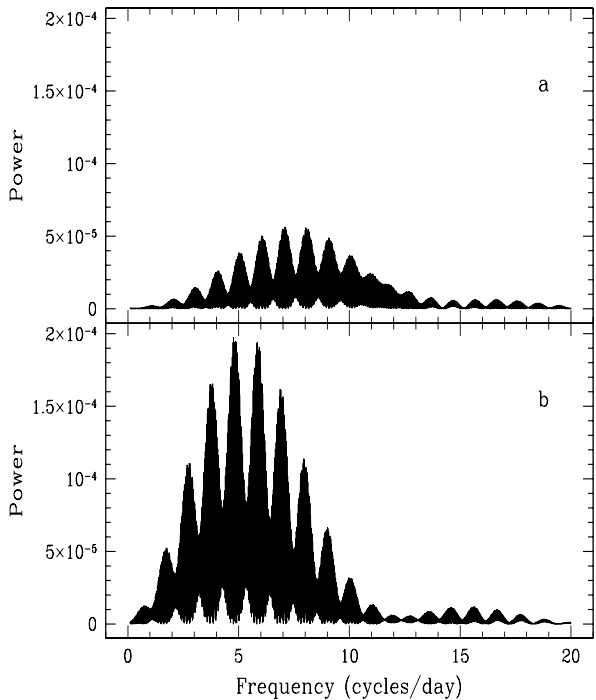,height=11cm,width=8.5cm,angle=0}
\caption[]{DFT power spectra of the $V$ data set after the subtraction of 
{\bf a} the 3$^{\rm h}$.46 and {\bf b} of the 4$^{\rm h}$.00 periodicities, 
respectively. While in the former case the power of the peaks decreases, in 
the latter it doubles with respect to Fig. 1a}
\end{figure}

\begin{figure}
\psfig{file=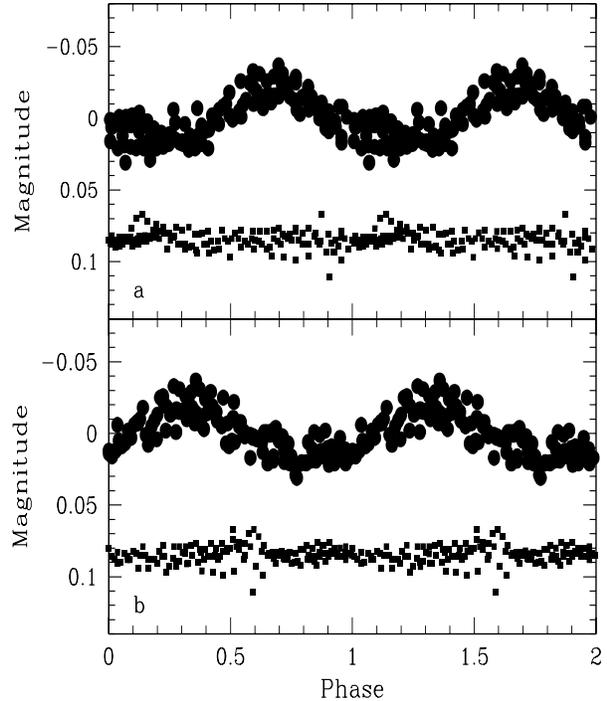,height=11cm,width=8.5cm,angle=0}
\caption[]{$V$ lightcurves (filled dots) folded with {\bf a} 
$P_{\rm orb}=3^{\rm h}.46$ and {\bf b} $P_{\rm orb}=4^{\rm h}.00$. Data points 
have been rescaled to a common zero magnitude level. Phases are arbitrarily 
referred to JD = 0.00. In both panels the mean of the internal differences 
among the comparison field stars is reported (filled squares), folded with the 
two periodicities and shifted of $-$0.18 mag for clarity}
\end{figure}

All the images taken at La Silla under photometric conditions (seeing 
$\la$1".5) seem to reveal around V1101 Aql the presence of an asymmetric 
nebulosity, best seen in $V$ (Fig. 4a) rather than in $B$ (Fig. 4b).
This feature was not detected in the frames obtained on July 1997 frames due to
poor seeing conditions.

To better disentangle the faint nebulosity from the background, we have 
designed, through DAOPHOT II, the average PSF using nearby field stars, and 
then we have subtracted it to the image of V1101 Aql. The result is presented
in Fig. 5a,b: an underlying bow--shaped nebulosity is visible in $V$ (Fig. 5a)
and in $B$ (Fig. 5b). 

\begin{figure*}
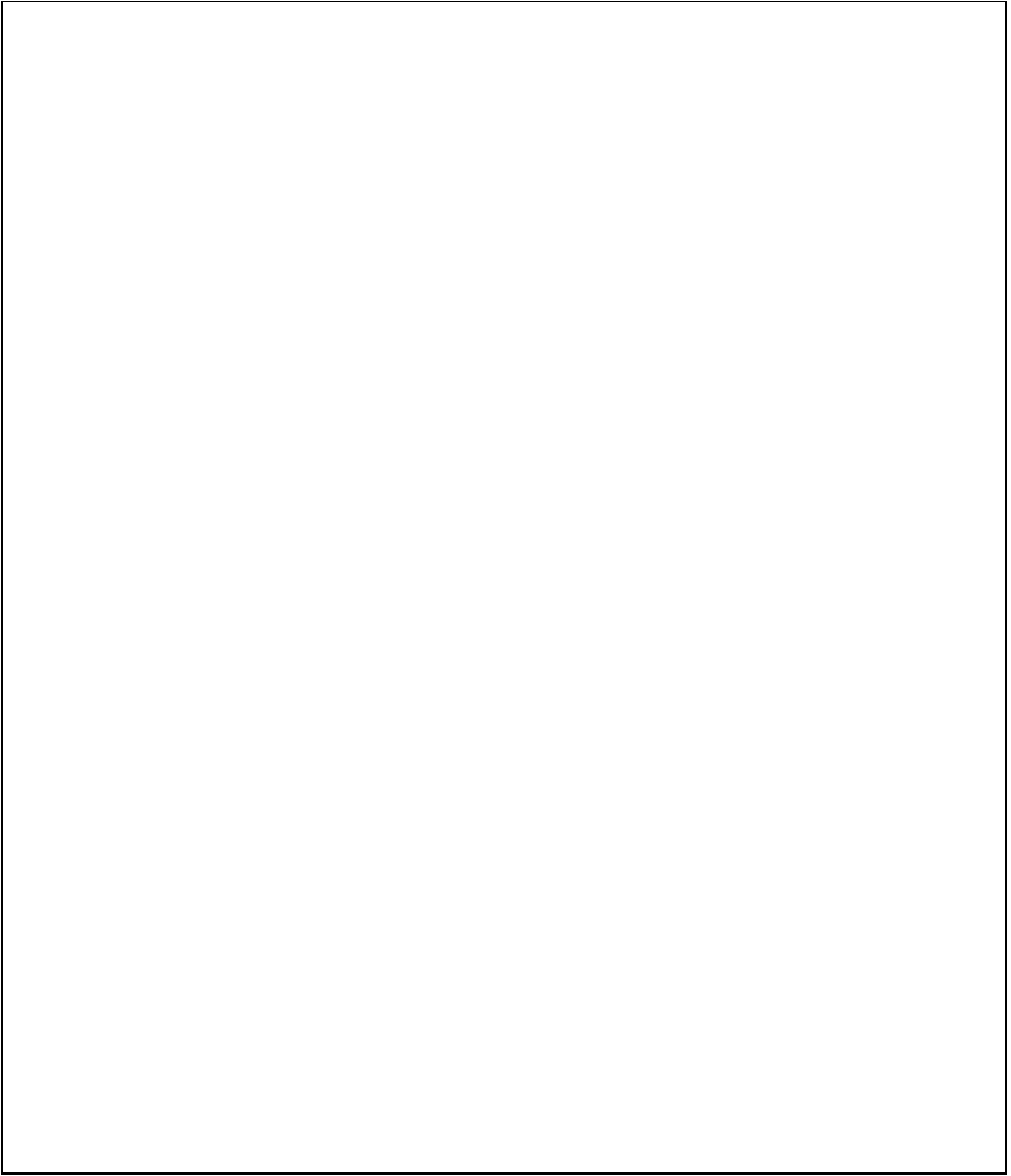

\picplace{210mm}
\caption[]{{\bf a} $V$ and {\bf b} $B$ images of the asymmetric nebulosity
around V1101 Aql (exposure times: 2 and 3 minutes, respectively). The field is 
0'.75$\times$0'.75 ; north is at top and east is on the right}
\end{figure*}

\begin{figure*}
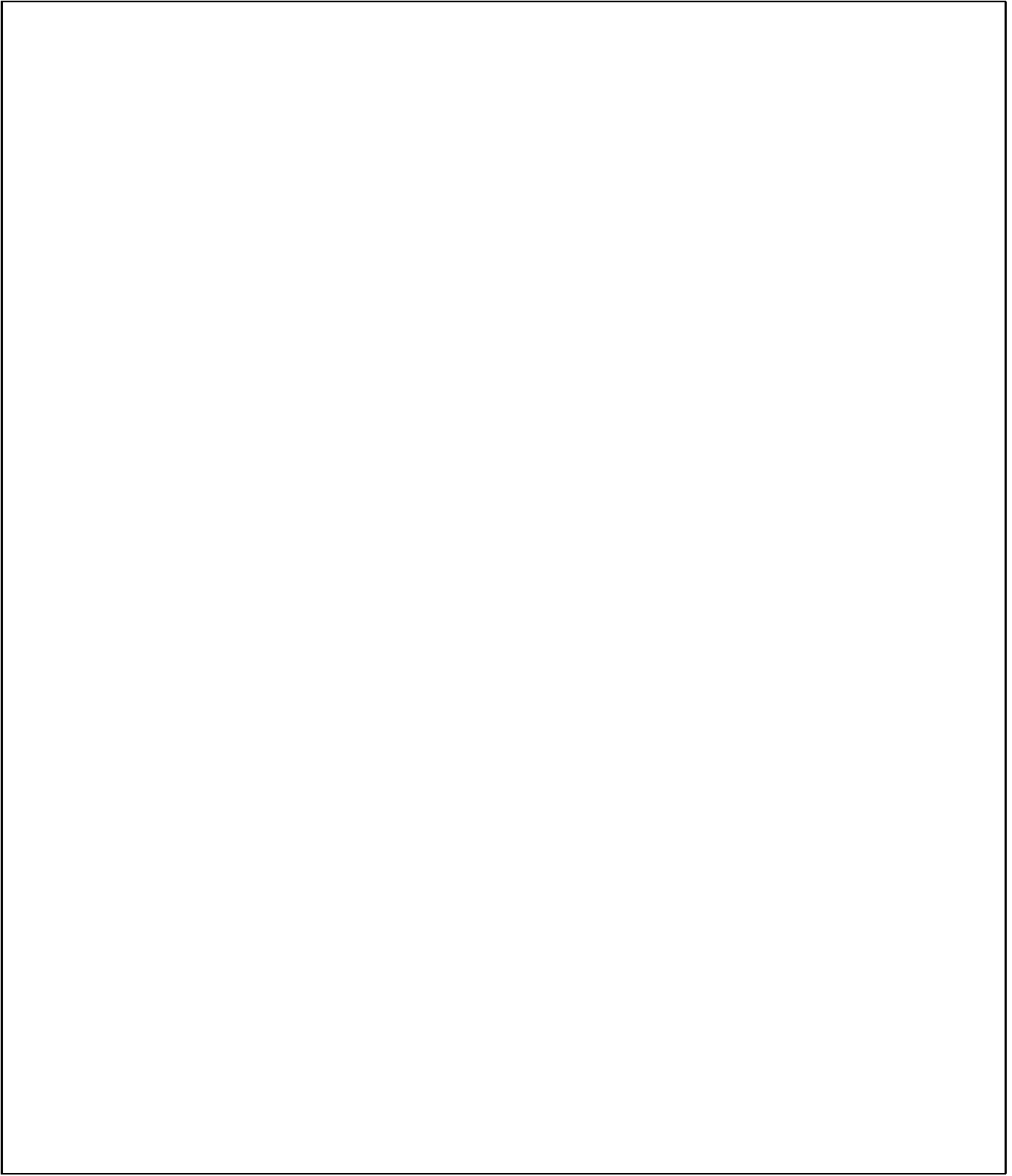

\picplace{210mm}
\caption[]{The same as Fig. 4, but after the subtraction of V1101 Aql. The
nebulosity appears brighter and more characterized {\bf a} in the $V$ band than
{\bf b} in the $B$}
\end{figure*}

\subsection{Spectroscopy}

The spectrum secured on June 9, 1996, is shown in Fig. 6. We
can notice the Balmer lines in absorption with a small emission core and, 
perhaps, the presence of a noisy emission of He {\sc ii} at 4686 \AA. 
Fluxes and EW's of these lines are reported in
Table 2. H$_\alpha$ is outside the spectral range as well as
the absorption lines and bands of the secondary, possibly present at longer
wavelengths.

The NaD interstellar absorption at 5890 \AA~is also present. This line might be 
possibly contamined by the presence of He {\sc i} $\lambda$5876 in absorption; 
a double gaussian fit yields EW$_{\rm NaD} = 1.2$ \AA, which corresponds,
according to the relation by Barbon et al. (1990), to a $E(B-V)\sim0.3$ mag.
This value is higher than that ($\approx$0.1 mag) found by 
Pastukhova \& Shugarov (1994) using the ($U-B$)/($B-V$) color ratio.


\begin{figure}
\psfig{file=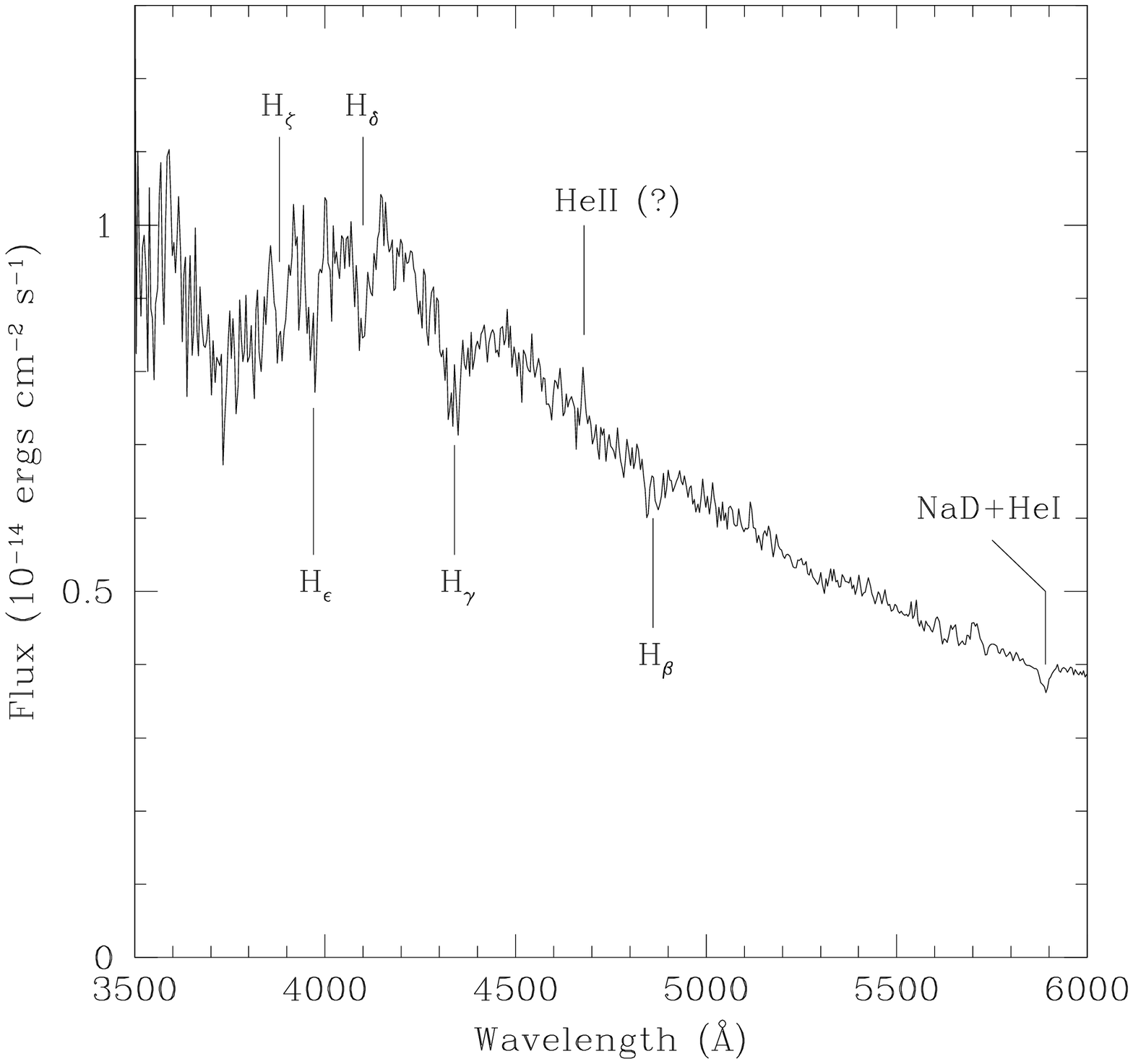,height=7cm,width=8.5cm,angle=0}
\caption[]{Spectrum of V1101 Aql acquired on June 9, 1996. No correction
for interstellar absorption has been applied. The major features indicated here
are described in the text}
\end{figure}

\begin{table}
\caption[]{Fluxes (in units of 10$^{-15}$ erg cm$^{-2}$ s$^{-1}$) and EW's 
(in \AA) of the lines detected in the spectrum of Fig. 6. If an emission core 
is present, emission and absorption fluxes and EW's are both reported}
\begin{center}
\begin{tabular}{cc|cc|c}
\noalign{\smallskip}
\hline
\noalign{\smallskip}
Spectral line & \multicolumn{2}{c}{Flux} & \multicolumn{2}{c}{EW} \\
\noalign{\smallskip}
\hline
\noalign{\smallskip}
 & em. & abs. & em. & abs.\\
\noalign{\smallskip}
\hline
\noalign{\smallskip}
\multicolumn{1}{l}{H$_\beta$} & 7.2 & 18.7 & 1.2 & 2.9 \\
\multicolumn{1}{l}{H$_\gamma$} & 11 & 68.9 & 1.6 & 7.9 \\
\multicolumn{1}{l}{H$_\delta$} & --- & 67.9 & --- & 6.8 \\
\multicolumn{1}{l}{H$_\epsilon$} & 3.1 & 58.0 & 0.38 & 5.9 \\
\multicolumn{1}{l}{H$_\zeta$} & 4.9 & 81.4 & 0.62 & 7.9 \\
\multicolumn{1}{l}{He {\sc i}} $\lambda$5876 & --- & 1.8 & --- & 0.46 \\
\noalign{\smallskip}
\hline
\end{tabular}
\end{center}
\end{table}

\section{Discussion}

\subsection{The orbital period and the distance to V1101 Aql}

We can interpret the modulation in the $V$ lightcurve of V1101 Aql as due to 
the orbital motion of the secondary star around the hot WD
(see Warner 1995 and references therein). The UV illumination
from the disk and the WD heats the inner face of the secondary and makes it
brighter than the other side. This, combined with the orbital motion, produces
a sinusoidal lightcurve with the maximum in correspondence of the superior
conjunction of the secondary. The small amplitude of the modulation
also indicates that eclipses are
absent, and thus that the inclination of the system must be small.

The periodic variation of $\sim$3$^{\rm h}$.5 is consistent with the mean
orbital period of Z Cam stars, which is always more than 3 hours (Warner 1995).
DNe with these periods have early--mid M type secondaries (Ritter \& Kolb 1995)
with masses around 0.3--0.4 $M_\odot$.

This orbital period determination can also lead us to attempt an estimate of 
the absolute magnitude M$_V$ of the system and hence its distance. 
According to Warner (1995), 
Z Cam DNe with orbital periods around 3$^{\rm h}$.5 should have M$_V\sim6.5
\pm0.3$ (1$\sigma$) at standstill. If V1101 Aql has, at standstill, 
$V\sim14.7$, (we assume the magnitude at quiescence to be $V\sim17$; see 
Pastukhova \& Shugarov 1994)
we derive, after taking into account the interstellar absorption 
in the $V$ band, a distance of $\sim300\pm50$ pc. This value is about one 
half the previous estimate given by Pastukhova \& Shugarov (1994) and, together
with the galactic latitude $b^{\rm II}=-10^{\circ}.16$, implies a height on the 
galactic plane $z\sim55$ pc.


\subsection{The spectrum and the nebulosity}

The spectrum presented in Fig. 6 shows the presence of absorption Balmer lines 
filled in with emissions, thus supporting, in agreement with Pastukhova \& 
Shugarov (1994), that this system might be a CV.

Concerning the asymmetric nebulosity around V1101 Aql, we note that it is
more visible in $V$ than in $B$ (see Fig. 5a,b). Actually, simple aperture 
photometry gives $V=18.64\pm0.05$ and $B-V=0.6\pm0.1$.
The total angular size of the nebulosity is $\sim$6 arcsecs; this, together 
with our distance estimate, leads to a linear size of $\approx$10$^{16}$ cm.


\section{Conclusions}

We observed long--term variations of V1101 Aql in $V$ magnitude, very likely 
triggered by disk activity.
We have detected $P=3^{\rm h}$.46 as possible orbital period of V1101 Aql, 
with an important alias at 4$^{\rm h}$.00.

The secondary of a DN with such an orbital period
should have M$_V\sim10$ (Warner 1995).
This means that the $V$ emission of the UV--heated 
secondary star is about 25 times lower than that of the whole system; 
i.e., any variation coming from the secondary could modify the total $V$ 
luminosity of the system by a factor $\la$1/25 at most, corresponding to a 
full amplitude fluctuation of $\la$0.04 mag.
This is smaller than the amplitude ($\sim$0.1 mag) of the modulation in Fig. 
3a,b, then indicating an extra contribution of $\sim$0.06 mag due to the UV 
heating of the inner face of the secondary.


The discovery of a Z Cam--type DN with an orbital period just above to the  
3--hr upper limit of the Period Gap of CVs is consistent with the 
presence of a sharp `luminosity bump', in the period--absolute magnitude
plane of CVs between 3 and 3.5 hr as suggested by Zangrilli et al. (1997).


Our results then, complemented with data gathered from literature, allow us to 
set the standstill magnitude of this Z Cam star close to $V\sim14.7$ and to
estimate for this system a distance of about 300 pc.

The spectrum, taken during standstill, seems to confirm the
cataclysmic binary nature of V1101 Aql, though the discovery of
an asymmetric nebulosity, which seems to 
be associated to the object, does not rule out the possibility, already pointed 
out by Downes et al. (1995), that V1101 Aql could be a Herbig Ae/Be star
rather than a CV. In this case, the presence of an associate nebulosity around
the object is fully consistent with such a classification.
However, due to the relatively poor seeing conditions ($\sim$1".5), we cannot
rule out that the image of V1101 Aql has been contamined by a background object.

\begin{acknowledgements}

The authors are indebted to the referee F. Verbunt for his valuable suggestions.
This research made use of the SIMBAD data base, operated by Centre de
Donn\'ees Stellaires in Strasbourg, France. We thank H.W. D\"urbeck for
important comments. NM thanks A. Bianchini and R. Claudi 
for having given him part of their observational time for the acquisition of 
the spectrum presented here, and A. Aloisi for useful suggestions on the 
advanced use of DAOPHOT II.

\end{acknowledgements}

\end{document}